 \documentstyle[12pt]{article}
 \newcommand{\eq}{\begin{equation}}
 \newcommand{\en}{\end{equation}}
 \newcommand{\eqn}{\begin{eqnarray}}
 \newcommand{\enn}{\end{eqnarray}}
 \newcommand{\nn}{\nonumber \\}

 \newcommand{\barr}{\begin{array}}
 \newcommand{\earr}{\end{array}}

 \newcommand{\sca}{superconformal algebra}
 \newcommand{\NP}{{\it Nucl. Phys.}}
 \newcommand{\PL}{{\it Phys.Lett.}}
 \newcommand{\CMP}{{\it Comm. Math. Phys.}}

 \newcommand{\A}{\alpha}
 \newcommand{\B}{\beta}

 \newcommand{\AP}{\alpha^{\prime}}
 \newcommand{\BP}{\beta^{\prime}}
 \newcommand{\APP}{\alpha^{\prime \prime}}
 \newcommand{\BPP}{\beta^{\prime \prime}}

\begin{document}

\title{  Extended Superconformal Symmetry , \\
 Freudenthal Triple Systems and \\
 Gauged WZW Models \thanks{Invited talk presented at G\"{u}rsey Memorial
Conference I , Istanbul, T\"{u}rkiye (June 6-10, 1994)} }

\author{Murat G\"{u}naydin \thanks{ E-mail address: murat@phys.psu.edu} \\
Physics Department, 104 Davey Lab. \\
Penn State University ,University Park, PA 16802}
\maketitle

\begin{flushleft}
{\bf Abstract:} We review the  construction of  extended  ( $N=2$ and $N=4$ )
superconformal algebras over triple systems and the gauged WZW models invariant
under them.
 The $N=2$ superconformal algebras (SCA) realized
over Freudenthal triple systems (FTS) admit extension to
``maximal''  $N=4$ SCA's with $SU(2) \times SU(2) \times U(1)$ symmetry.
A detailed  study of the construction and classification
of $N=2$ and $N=4$ SCA's over
Freudenthal triple
systems is given.
We conclude with
 a  study and classification of gauged WZW
 models with $N=4$ superconformal symmetry.

\end{flushleft}

 \section{Introduction}
 \setcounter{equation}{0}
It is a singular honor for me to give a talk in the First of the G\"{u}rsey
Memorial Conferences which are to be held biannually. I regard Feza G\"{u}rsey
as my mentor and a role model I try to emulate. He has had a great influence
on my style of physics.

 In this talk I will discuss extended superconformal
algebras and their connection with triple systems, in particular the
Freudenthal Triple Systems. I will also discuss a class of Lagrangian field
theories,
namely the gauged WZW models, that are invariant under extended superconformal
groups. I learned about Freudenthal triple systems more than twenty years ago
 when I was working with
Feza on the questions of physical implications of extending the underlying
number system of quantum mechanics from complex numbers to octonions and
what possible role exceptional groups may play in physics.

Infinite conformal algebra in two dimensions and its
supersymmetric
extensions have been studied extensively in recent years.
They underlie string and superstring theories as their local
gauge symmetries\cite{GSW}.
The classical vacua of string theories are described
 by conformal field theories. For example,
 the heterotic string vacua with $N = 1$ space-time
 supersymmetry in four dimensions are described by  "internal" $N = 2$
 superconformal field theories with  central charge $c = 9$
 \cite{BDFM,N=2,LVW,KS89,DLP,IB}.
 The $N = 2$ space-time supersymmetric vacua of the heterotic string
 are described by an internal superconformal field theory with four
 supersymmetries \cite{NS}.
 The extended superconformal algebras
 have important applications to
integrable systems \cite{IS} and to
topological field theories in two dimensions as well
\cite{TFT}.
The conformal and superconformal
algebras  have been studied in detail via
 the coset space method of Goddard, Kent and Olive (GKO)
 \cite{GKO} which is a generalization
of Sugawara-Sommerfield construction of the Virasoro algebra in terms of
bilinears of the generators of a current algebra \cite{SS}.
In the first part of my talk I will review a novel
 ternary
algebraic approach to the construction and study of extended superconformal
algebras.
The construction of N=2 SCA's over Jordan Triple
Systems (JTS) was given in \cite{MG91}.
This construction was generalized to
 the realization of N=2 SCA's over more general Kantor
triple systems (KTS) in \cite{GH91,GH92}.
 The coset spaces associated with the KTS's
are, in general, not symmetric spaces.
 For a particular subclass of Kantor triple systems, namely the
 Freudenthal triple systems (FTS) the N=2 SCA's admit extensions to ``maximal''
$N=4$ SCA's with the gauge group $ SU(2) \times SU(2) \times U(1) $
 \cite{STV}.
The construction of $N=2$ and $N=4$ superconformal algebras
over FTS's and their classification will be discussed in detail
following \cite{GH91,GH92,MG92,MG93a}.
The last part of my talk is devoted to the study of gauged WZW models
that are invariant under $N=4$ superconformal symmetry \cite{MG93b}.

 \section{Construction of Extended Superconformal Algebras over Triple
  Systems}

 \setcounter{equation}{0}

  The realization of $N=2$ \sca s over hermitian Jordan triple
 systems given in \cite{MG91} is equivalent to their realization over hermitian
symmetric spaces
 of Lie groups a la Kazama and Suzuki \cite{KS89} which can be compact or
 non-compact .
 The connection between hermitian symmetric spaces
 and hermitian Jordan triple systems arises as follows.
 If $G/H$ is a hermitian
 symmetric space then the Lie algebra $g$ of $G$ can be given a 3-graded
decomposition with respect to the Lie algebra $g^{0}$ of $H$:
 \begin{equation}
 g = g^{-1} \oplus g^{0} \oplus g^{+1}
 \end{equation}
 where $ \oplus $ denotes vector space direct sum and $g^{0}$ is a
 subalgebra of maximal rank. We have the formal commutation relations of the
  elements of various grade subspaces
 \begin{equation}
 [g^{m},g^{n}] \subseteq g^{m+n} \; ; m,n=-1,0,1
 \end{equation}
 where $ g^{m+n} = 0 $ if $|m+n|>1$.  The Tits-Kantor-Koecher (TKK)
   construction \cite{TKK} of the Lie algebra $g$ establishes a  mapping
 between the grade $+1$ subspace of $g$ and the underlying  Hermitian JTS $V$.

 The exceptional Lie algebras $G_{2}, F_{4}$ and
 $E_{8}$ do not admit a TKK type construction. A generalization
 of the TKK construction to more general triple systems was given
 by Kantor \cite{IK}. All finite dimensional simple Lie algebras
 admit a construction over these generalized triple systems.
 The Kantor's  construction  of
 Lie algebras was generalized to a unified construction of Lie algebras
 and Lie superalgebras in \cite{BG79} which we shall review briefly restricting
our discussion to Lie algebras only.

 This more general construction starts from the fact that
 every simple Lie algebra $g$ with the exception of $SU(2)$ admits  a 5-grading
(Kantor structure) with
 respect to some
 subalgebra $g^{0}$ of maximal rank \cite{IK,BG79}:
 \begin{equation}
 g = g^{-2} \oplus g^{-1} \oplus g^{0} \oplus g^{+1} \oplus g^{+2}
 \end{equation}
 One labels the elements  of $g^{+1}$ subspace with
 the elements of a vector space
  $V$ \cite{IK,BG79}:
 \begin{equation}
 U_{a} \in g^{+1} \Longleftrightarrow a \in V
 \end{equation}
  $g$ admits a conjugation  under which the grade
 $+m$ subspace  gets mapped into the grade $-m$ subspace, which allows one
 to label the elements of the grade $-1$
 subspace by the elements of $V$ as well :
 \begin{equation}
 U^{a} \equiv U_{a}^{\dagger} \in g^{-1}
 \Longleftrightarrow U_{a} \in g^{+1}
 \end{equation}
 One defines the commutators of $U_{a}$ and $U^{b}$ as
 \begin{equation}
 \begin{array}{l}
 {[}U_{a} , U^{b}{]} = S_{a}^{b} \in g^{0}  \\
 {[}U_{a} , U_{b}{]} = K_{ab} \in g^{+2}               \\
 {[}U^{a} , U^{b}{]} = K^{ab} \in g^{-2}             \\
 {[}S_{a}^{b} , U_{c}{]} = U_{(abc)} \in g^{+1}

 \end{array}
 \end{equation}
 where $(abc)$ is the triple  product under which the
 elements of $V$ close. The remaining non-vanishing commutators of $g$ can
 all be expressed in terms of the triple product $(abc)$:
 \begin{equation}
 \begin{array}{l}

 {[}S_{a}^{b} , U^{c}{]} = - U^{(bac)}           \\
 {[}K_{ab} , U^{c}{]} = U_{(acb)} - U_{(bca)}  \\
 {[}K^{ab} , U_{c}{]} = - U^{(bca)} + U^{(acb)}\\
 {[}S_{a}^{b} , S_{c}^{d}{]} = S_{(abc)}^{d} - S_{c}^{(bad)}      \\
 {[}S_{a}^{b} , K_{cd}{]} = K_{(abc)d} + K_{c(abd)} \\
 {[}S_{a}^{b} , K^{cd}{]} = - K^{(bac)d} - K^{c(bad)} \\
 {[}K_{ab} , K^{cd}{]} = S_{(acb)}^{d} - S_{(bca)}^{d} - S_{(adb)}^{c} +
 S_{(bda)}^{c}
 \end{array}
 \end{equation}
 The Jacobi identities of $g$ follow from
  the defining  identities  of a Kantor triple system (KTS) \cite{IK,BG79}
 \begin{equation}
 \begin{array}{l}
 (ab(cdx)) - (cd(abx)) - (a(dcb)x) + ((cda)bx) = 0
 \end{array}
 \end{equation}
  \begin{equation}
 \begin{array}{l}
 \{ (ax(cbd)) - ((cbd)xa) + (ab(cxd)) + (c(bax)d) \} - \{ c
 \leftrightarrow d \} = 0
 \end{array}
 \end{equation}
 In general a given simple Lie algebra can be constructed in
 several different ways by the above method
 corresponding to different choices of the subalgebra $g^{0}$ and
 different KTS's. Note that the second defining identity of a KTS is trivially
 satisfied by a JTS for which the grade $\pm 2$ subspaces vanish.

Consider now
 the affine Lie algebra $\hat{g}$ defined by
  the Lie algebra $g$ constructed over a KTS $V$.
 The commutation relations
 $\hat{g}$ can be written in the form of operator products as follows:
 \begin{equation}
 \begin{array}{lll}

 U_{a}(z) U^{b}(w) & = &
 \frac{k \delta_{a}^{b}}{(z - w)^{2}} + \frac
 {S_{a}^{b}(w)}{(z - w)} + \cdots \\
 U^{a}(z) U_{b}(w) & = &
 \frac{k \delta_{a}^{b}}{(z - w)^{2}} - \frac
 {S_{a}^{b}(w)}{(z - w)} + \cdots \\
 U_{a}(z) U_{b}(w) & = &
 \frac{K_{ab}(w)}{(z - w)} + \cdots \\
 S_{a}^{b}(z) U_{c}(w) & = &
 \frac{U_{(abc)}(w)}{(z-w)} + \cdots \\
 S_{a}^{b}(z) U^{c}(w) & = &
 \frac{-U^{(bac)}(w)}{(z-w)} + \cdots \\
 S_{a}^{b}(z) S_{c}^{d}(w) & = &
 \frac{k \Sigma_{ac}^{bd}}{(z-w)^{2}}
  - \frac{1}{(z-w)} (S_{(abc)}^{d} - S_{c}^{(bad)})(w) + \cdots \\
 S_{a}^{b}(z) K_{cd}(w) & =  &
  \frac{1}{(z-w)} (K_{(abc)d} + K_{c(abd)})(w) + \cdots \\
 S_{a}^{b}(z) K^{cd}(w) & = &
  \frac{-1}{(z-w)} (K^{(bac)d} + K^{c(bad)})(w) + \cdots \\
 K_{ab}(z) K^{cd}(w) & = & \frac{k \Omega_{ab}^{cd}}{(z-w)^{2}} +
  \frac{1}{(z-w)} (S_{(acb)}^{d} - S_{(bca)}^{d}
  - S_{(adb)}^{c} + S_{(bda)}^{c})(w) + \cdots \\
 K_{ab}(z) U^{c}(w) & = & \frac{1}{(z-w)} (U_{(acb)} - U_{(bca)})(w)
  + \cdots  \\
 K^{ab}(z) U_{c}(w) & = & \frac{1}{(z-w)} (U^{(acb)} - U^{(bca)})(w)
  + \cdots

 \end{array}
 \end{equation}
 where $\Sigma_{ab}^{cd} $ are the structure constants of the underlying KTS
 $ V$:
 \begin{equation}
 U_{(abc)} = \Sigma_{ac}^{bd} U_{d}
 \end{equation}
 and the tensor $\Omega$ is defined as
 \begin{equation}
   \Omega_{ab}^{cd}  = \Sigma_{ab}^{cd}  -  \Sigma_{ab}^{dc}
 \end{equation}
 We choose a basis
   $ V$ such that

 \begin{equation}
 \begin{array}{lll}
 \Omega_{ab}^{ca} & = & \frac{2d}{D} (\check{g} -
 \check{s})\delta_{b}^{c}
  \end{array}
 \end{equation}
 where $\check{g}$  and $\check{s}$ denote the dual coxeter numbers  of
 the  Lie
 algebra $g$ and its subalgebra $s$ generated by the elements of the grade
 $\pm2$ subspaces and their commutant $[g^{-2},g^{+2}]$, respectively.
 $D$ is the dimension of the triple system $V$
 and  $d$ is the dimension of the grade $+2$ subspace.

 Furthermore one  introduces fermion fields $ \psi_{a} (\psi^{a})$ and
 $ \psi_{ab} (\psi^{ab}) $ corresponding to the grade $ +1 (-1) $
 and $+2 (-2)$ subspaces of $g$, respectively.
  They are normalized such that
 \begin{equation}
 \begin{array}{l}

 \psi_{a}(z) \psi^{b}(w) = \frac{\delta_{a}^{b}}{(z - w)} + \cdots  \\
 \psi_{ab}(z) \psi^{cd}(w) = \frac{\Omega_{ab}^{dc}}{(z - w)} + \cdots

 \end{array}
 \end{equation}
 The operators of conformal dimension $3/2$
  defined  by the following expressions:
 \begin{equation}
 \begin{array}{l}

 G(z) = \sqrt{\frac{2}{k+\check{g}}} \{ U_{a} \psi^{a} +
 \frac{1}{2(\check{g}-\check{s})}
 K_{ab} \psi^{ab} - \frac{1}{2} \psi^{a} \psi^{b} \psi_{ab} \}(z) \\
 \bar{G}(z) = \sqrt{\frac{2}{k+\check{g}}} \{ U^{a} \psi_{a} -
 \frac{1}{2(\check{g}-\check{s})}
 K^{ab} \psi_{ab} + \frac{1}{2} \psi^{ab} \psi_{a} \psi_{b} \}(z)
 \end{array}
 \end{equation}
  generate an N = 2 superconformal algebra \cite{GH91,GH92} with the Virasoro
 generator
 \begin{equation}
 \begin{array}{lll}
 T(z) & = & \frac{1}{k+\check{g}} \{ \frac{1}{2}(U_{a}U^{a} +
 U^{a}U_{a})
 + \frac{1}{4(\check{g}-\check{s})}(K_{ab}K^{ba} + K^{ab}K_{ba})   \\
      &   & +\frac{k}{2}( \partial \psi^{a}  \psi_{a}+
 \partial \psi_{a} \psi^{a} )  + \frac{1}{2} \Omega_{ac}^{cb} ( \partial
 \psi^{a}  \psi_{b} + \partial \psi_{b} \psi^{a})  \\
 & & + \frac{(k+\check{g}-\check{s})}{4(\check{g}-\check{s})} ( \partial
 \psi^{ab}\psi_{ab} +
  \partial \psi_{ab} \psi^{ab} )
 + S_{a}^{b} \psi^{a} \psi_{b}   \\
 & &  + \frac{1}{(\check{g}-\check{s})} S_{a}^{b} \psi^{ac} \psi_{bc} +
 \psi_{ab} \psi^{ac} \psi^{b}
 \psi_{c} +
 \frac{1}{4}\Omega_{ab}^{cd} \psi^{a} \psi^{b} \psi_{c} \psi_{d} \}(z)
 \end{array}
 \end{equation}
 and the $U(1)$ current
 \\
 \begin{equation}
 \begin{array}{lll}
 J(z)  & = & \frac{1}{k+\check{g}} \{ S_{a}^{a}  - \frac{1}{(\check{g} -
 \check{s})} \Omega^{cb}_{ca} S^{a}_{b}+ k \psi^{a}
  \psi_{a}   \\
 & & +\Omega^{bc}_{ab} \psi^{a} \psi_{c} + \frac{(k-
 \check{g}+\check{s})}{2(\check{g}-\check{s})} \psi^{ab} \psi_{ab} \} (z)
 \end{array}
 \end{equation}
  Its central charge turns out to be :
 \begin{equation}
  c = \frac{3}{(k+\check{g})} \{ kD + \frac{k}{4(\check{g}-\check{s})^2}
 \Omega^{cd}_{ab} \Omega^{ab}_{cd} + \frac{1}{2}
 \Omega^{ba}_{ab}\}
 \end{equation}

 This realization of N=2 SCA's  is equivalent
 to their realization over the coset  spaces
 $G/H$  where $G$ and $H$ are the groups generated by $g$ and $g^{0}$,
 respectively. These spaces are, in general, not symmetric spaces.
  For further details and a complete classification of the
 $N=2$ SCA's constructed over KTS's we refer to \cite{GH91,GH92}.

 \section{ Construction of $N = 2$ Subalgebras of Maximal $N = 4$
 Superconformal
 Algebras over Freudenthal Triple Systems}
 \setcounter{equation}{0}

  For a very special subclass of Kantor triple systems,
 namely the Freudenthal triple systems (FTS), the $N=2$ superconformal algebras
as constructed in the previous section
 can be extended to  the maximal $N=4$  \sca s \cite{GH91,GH92}.
 Freudenthal introduced these triple systems in his study of the geometries
 associated with exceptional groups \cite{HF}.
 Kantor and Skopets classified FTS's and showed that there is a one-to-one
correspondence between simple
 Lie algebras and simple FTS's with a non-degenerate bilinear form \cite{KS82}.
  The Freudenthal triple product $(abc)$ can  be written in the form:
 \eq
 (abc)= \{abc\}-(c,b)a-(c,a)b-(a,b)c
 \en
 where $\{abc\}$ is completely symmetric in its arguments and $( , )$
  is a skew-symmetric bilinear form defined over the FTS.
 For  a simple
 Lie algebra $g$ constructed over  a FTS the grade $\pm 2$ subspaces become one
 dimensional as a consequence of the identity
 \eq
 (abc)-(cba)= 2(a,c)b
 \en
 The
 realization of   $N=2$ superconformal algebra  over it corresponds to
  the coset space  $G/H_{0}\times U(1)$ where $H_{0}$ is such that
 $G/H_{0}\times SU(2)$ is  the unique quaternionic symmetric space  of $G$
 .

 Let $g,h_{0}$ be the Lie algebras of $G$ and $H_{0}$ listed above,
  respectively.
 The 5-graded
 structure of $g$
 \eq
 g = g^{-2} \oplus g^{-1} \oplus g^{0} \oplus g^{+1} \oplus g^{+2}
 \en
 is such that $g^{0}=h_{0} \oplus K_{3}$ where $K_{3}$ is the generator of the
$U(1)$ factor and the elements $K_{ab}$ and $K^{ab}$ of
 grade $\pm 2$ subspaces  can be written in the form
 \eqn
 K_{ab} &=& \Omega_{ab} K_{+} \nn
 K^{ab} &=& \Omega^{ab} K^{+} \nonumber
 \enn
 where $\Omega_{ab}$ is a symplectic form defined over the FTS. \footnote{ In
 a given basis $e_{a}$ of the FTS one can set $\Omega_{ab}= (e_{a},e_{b})$.}
 The tensor $\Omega^{ab}$ is the inverse of
 $\Omega_{ab}$ and
 \eqn
 \Omega_{ab} \Omega^{bc}& =& \delta_{a}^{c} \nn
 \Omega_{ab}^{\dagger}& =& \Omega^{ba} =-\Omega^{ab}
 \enn
 The elements $K_{+}$ and $K^{+}$ are hermitian conjugates of each other
 and can be written as
 \eqn
 K_{+}&=&K_{1}+iK_{2} \in g^{+2} \nn
 K^{+}& \equiv & K_{-}= K_{1}-iK_{2} \in g^{-2}
 \enn
  There is a universal relation between  the dual Coxeter number $\check{g}$ of
$G$ and
 the dimension $D$ of the underlying FTS  :
 \eq
 D=2(\check{g}-2)
 \en
 The generators $K_{-}, K_{+}$ and $K_{3}$ form an $SU(2)$ subalgebra of $g$.
 \eqn
 {[K_{+},K_{-}]}& =&2K_{3} \nn
 {[K_{3}, K_{\pm}]}& =& \pm K_{\pm}
 \enn
 The commutation relations of the U's can now be written in the form
 \eqn
 {[U_{a},U_{b}]}& =& \Omega_{ab} K_{+} \nn
 {[U^{a}, U^{b}]}& =& \Omega^{ab} K_{-} \nn
 {[U_{a}, U^{b}]}& =& S_{a}^{b}
 \enn

  $\Omega_{ab}$ is an invariant tensor of $H_{0}$ and $S_{a}^{b}$ are the
 generators of the subgroup $H_{0}\times U(1)$.
 The trace component of $S_{a}^{b}$ gives the $U(1)$ generator
 \eq
 K_{3} =\frac{1}{2(\check{g}-2)} S_{a}^{a}
 \en
 Hence we have the decomposition
 \eq
 S_{a}^{b} = H_{a}^{b} + \delta_{a}^{b} K_{3}
 \en
 where $H_{a}^{b} = S_{a}^{b} - \frac{1}{D} \delta_{a}^{b} S_{c}^{c}$ are the
generators of the subgroup $H_{0}$. Note that $H_{a}^{b}$ commutes with $K_{3},
K_{+} $ and $K_{-}$.
 The other non-vanishing commutators of $g$ are
 \eqn
 {[K_{+}, U^{a}]}& =& \Omega^{ab} U_{b} \nn
 {[K_{-}, U_{a}]}& =& \Omega_{ab} U^{b} \nn
 {[K_{3}, U^{a}]}& =& -{\textstyle \frac{1}{2}} U^{a} \nn
 {[K_{3}, U_{a}]}& =& {\textstyle \frac{1}{2}} U_{a} \nn
 {[S_{a}^{b}, U_{c}]}& =& \Sigma_{ac}^{bd} U_{d} \nn
 {[S_{a}^{b}, U^{c}]}& =& -\Sigma_{ad}^{bc} U^{d} \nn
 {[S_{a}^{b}, S_{c}^{d}]}& =& \Sigma_{ac}^{be} S_{e}^{d} -\Sigma_{ae}^{bd}
S_{c}^{e}
 \enn
 where $\Sigma_{ab}^{cd}$ are the structure constants of the  FTS
 which in our normalization satisfy
 \eqn
 \Sigma_{ab}^{ac}& =& (\check{g} -2) \delta_{b}^{c} \nn
 \Sigma_{ab}^{bc}& =& (\check{g} -1) \delta_{a}^{c}  \nn
 \Sigma_{ab}^{cd} - \Sigma_{ab}^{dc}& =& \Omega_{ab} \Omega^{cd}
 \enn

 The complex Fermi fields associated with the grade $\pm 2$
  subspaces can be represented as
 \eqn
 \psi_{ab}(z) &=& \Omega_{ab} \psi_{+}(z) \nn
 \psi^{ab}(z) &=& \Omega^{ab} \psi^{+}(z)
 \enn

 where   $\psi_{+}(z)$ and $\psi^{+}(z)$
 satisfy \cite{GH91,GH92}:

 \begin{equation}
 \begin{array}{l}

 \psi_{+}(z) \psi^{+}(w) = \frac{1}{(z - w)} + \cdots

 \end{array}
 \end{equation}

 The supersymmetry generators  of the $N=2$ superconformal algebra
 simplify

 \begin{equation}
 \begin{array}{lll}

 G(z) & = & \sqrt{\frac{2}{k+ \check{g}}}
  \{ U_{a} \psi^{a} + K_{+} \psi^{+} -
 \frac{1}{2} \Omega_{ab} \psi^{a} \psi^{b} \psi_{+} \}(z) \\
 \end{array}
 \end{equation}
 \begin{equation}
 \begin{array}{lll}
 \bar{G}(z) & = & \sqrt{\frac{2}{k+ \check{g}}}
  \{ U^{a} \psi_{a} +
  K^{+} \psi_{+} -
  \frac{1}{2} \Omega^{ab} \psi_{a} \psi_{b} \psi^{+} \}(z)  \\
 \end{array}
 \end{equation}
 The Virasoro generator takes the form

 \begin{equation}
 \begin{array}{lll}
 T(z) & = & \frac{1}{k+ \check{g}}
 \{ \frac{1}{2}(U_{a}U^{a} + U^{a}U_{a})
 + \frac{1}{2}(K_{+}K^{+} + K^{+}K_{+})   \\
      &   & - \frac{k+1}{2}(\psi_{a} \partial \psi^{a} +
 \psi^{a} \partial \psi_{a})
 - \frac{1}{2}(k+ \check{g}-2)(\psi_{+} \partial \psi^{+} +
 \psi^{+} \partial \psi_{+})   \\
       &   & + H_{a}^{b} \psi^{a} \psi_{b} +
 K_{3}(\psi^{a} \psi_{a} +2 \psi^{+} \psi_{+}) +
	\psi_{+} \psi^{+} \psi^{a} \psi_{a} +
 \frac{1}{4}\Omega_{ab} \psi^{a} \psi^{b} \Omega^{cd}
  \psi_{c} \psi_{d} \}(z)
 \end{array}
 \end{equation}

 and the $U(1)$ current is given by

 \begin{equation}
 J(z) = \frac{1}{k+ \check{g}} \{ 2(\check{g} - 1) K_{3}
  + (k+1)\psi^{a}
  \psi_{a} + (k- \check{g}+2) \psi^{+} \psi_{+} \}(z)
 \end{equation}
 The central charge of the N = 2 SCA defined by
 a FTS is
 \begin{equation}
  c = \frac{6(k+1)(\check{g}-1)}{(k+ \check{g})}-3
 \end{equation}

 As stated earlier the above realization of $N=2$ SCA's corresponds to the
coset $G/H_{0}\times U(1)$ where
 the $U(1)$ generator $K_{3}$  determines the 5-graded structure of $g$
 \eq
 {[2K_{3}, g^{m}]} = m g^{m}
 \en
 where $g^{m}$ denotes the subspace of grade $m $ with
 $m=0,\pm1,\pm2 $.
 The Lie algebra $g$  can also be given a 5-graded
  structure with respect to $K_{1}$ as well as $K_{2}$.
 Therefore, one can realize the $N=2$ SCA equivalently over the coset $G/H_{0}
\times U(1)^{\prime} $
 or the coset $G/H_{0} \times U(1) ^{\prime\prime}$, where the generators of
$U(1)^{\prime}$ and
 $U(1)^{\prime\prime}$ are $K_{1}$ and $K_{2}$, respectively. The generators
 belonging to
  grade $\pm1$ and $\pm2$ subspaces
 with respect to $K_{1}$ are
 \eqn
 U_{a}^{\prime}& =&{\textstyle \frac{1}{\sqrt{2}}} ( U_{a} + \Omega_{ab}
U^{b})\nn
 U^{a^{\prime}}& =&{\textstyle \frac{1}{\sqrt{2}}} ( U^{a} - \Omega^{ab} U_{b}
) \nn
 K_{+}^{\prime}& =& i(K_{2}+iK_{3}) \nn
 K_{-}^{\prime}& =& -i (K_{2}-iK_{3})
 \enn
 They satisfy
 \eqn
 {[U_{a}^{\prime},U_{b}^{\prime} ]}& =& \Omega_{ab} K_{+}^{\prime} \nn
 {[U^{a^{\prime}},U^{b^{\prime}}]}&=& \Omega^{ab} K_{-}^{\prime} \nn
 {[K_{+}^{\prime}, U^{a^{\prime}} ]}& =& \Omega^{ab} U_{b}^{\prime} \nn
 {[K_{-}^{\prime}, U_{a}^{\prime}]}& = & \Omega_{ab} U^{b^{\prime}}
	\label{eq:UP}
 \enn
 Whereas the grade $\pm1$ and $\pm2$ subspaces with respect to $K_{2}$ are
 \eqn
 U_{a}^{\prime\prime}& =&{\textstyle \frac{1}{\sqrt{2}}} (U_{a} +i \Omega_{ab}
U^{b})\nn
 U^{a^{\prime\prime}}& =&{\textstyle \frac{1}{\sqrt{2}}} (U^{a} +i\Omega^{ab}
U_{b} ) \nn
 K_{+}^{\prime\prime}& =& -i(K_{3} +iK_{1}) \nn
 K_{-}^{\prime\prime}& =& i(K_{3} -iK_{1})
  \enn
 with analogous  commutation relations to (\ref{eq:UP}).

 For every simple Lie group $G$ (except for $SU(2)$) there exists a subgroup
 $H_{0} \times U(1)$
 , unique up to automorphisms corresponding to $SU(2)$ rotations, such that
 its Lie algebra $g$ has a 5-graded structure with respect
 to the subalgebra $h_{0} \oplus K$, where $K$ is the generator of
  the $U(1)$ subgroup that determines the 5-grading.
  This follows from the fact that there is a one-to-one
  correspondence between simple Lie algebras and simple FTS's with a
  non-degenerate bilinear form
 \cite{KS82}.

\section{The Construction of Maximal $N = 4$ Superconformal Algebras}

The $N=2$ \sca s constructed over FTS's admit extensions to maximal
$N=4$ \sca s of references \cite{STV}.
To achieve this one needs to introduce a $N=2$ ``matter
multiplet''  and define two additional supersymmetry generators  as well as
adding the matter contributions to the first two
 supersymmetry generators.
The  required currents of the matter multiplet turn out to be the $U(1)$
current generated by $K_{3}$ that gives the
5-graded structure of the Lie algebra $g$, and an additional $U(1)$ current
whose
generator $K_{0}$ commutes with $g$ together with the associated fermions
which we denote
as a complex fermion $\chi_{+}$ and its conjugate $\chi^{+}$. Then
the four supersymmetry  generators of the $N=4$ \sca ~ can be written as

\begin{equation}
\begin{array}{lll}

G^{+} \equiv \frac{1}{\sqrt{2}} (G_{1}+ i G_{2}) & = &
\sqrt{\frac{2}{k+\check{g}}}
\{ U_{a} \psi^{a} + K_{+} \psi^{+} +
 K_{3} \chi_{+}  \\
&   & - \frac{1}{2} \Omega_{ab} \psi^{a} \psi^{b} \psi_{+} -
\frac{1}{2} \psi^{a} \psi_{a} \chi_{+} - \psi^{+} \psi_{+} \chi_{+} \}
+ iZ \chi_{+}  \\
G^{-} \equiv \frac{1}{\sqrt{2}} (G_{1}- i G_{2}) & = &
\sqrt{\frac{2}{k+\check{g}}}
\{ U^{a} \psi_{a} + K^{+} \psi_{+}
+ K_{3} \chi_{+} \\
&  & - \frac{1}{2} \Omega^{ab} \psi_{a} \psi_{b} \psi^{+} -
\frac{1}{2}\psi^{a} \psi_{a} \chi^{+} - \psi^{+} \psi_{+} \chi^{+} \}
-i Z \chi^{+}  \\
G^{+K} \equiv \frac{1}{\sqrt{2}}(G_{3}+ i G_{4}) & = &
\sqrt{\frac{2}{k+\check{g}}}
\{ \Omega^{ab} U_{a} \psi_{b} + K_{+} \chi^{+} +
K_{3} \psi_{+}  \\
&   & + \frac{1}{2} \Omega^{ab} \psi_{a} \psi_{b} \chi_{+} +
\frac{1}{2} \psi^{a} \psi_{a} \psi_{+} - \chi^{+} \chi_{+} \psi_{+} \}
+ iZ \psi_{+}  \\
G^{-K} \equiv \frac{1}{\sqrt{2}}(G_{3}- i G_{4}) & = &
\sqrt{\frac{2}{k+\check{g}}}
\{ \Omega_{ab} U^{b} \psi^{a} + K^{+} \chi_{+} +
K_{3} \psi^{+}  \\
&   & + \frac{1}{2} \Omega_{ab} \psi^{a} \psi^{b} \chi^{+} +
\frac{1}{2} \psi^{a} \psi_{a} \psi^{+} - \chi^{+} \chi_{+} \psi^{+} \}
-i Z \psi^{+}  \\

\end{array}
\end{equation}

The Virasoro generator of the maximal $N=4$ \sca ~
${\cal A}_{\gamma}$  is given by
\begin{equation}
\begin{array}{lll}
T(z) & = & \frac{1}{2} \left[ Z^{2} -
 (\chi_{+} \partial \chi^{+} +
\chi^{+} \partial \chi_{+}) - (\psi_{+} \partial \psi^{+} +
\psi^{+} \partial \psi_{+}) \right](z)   \\
     &   & + \frac{1}{k+\check{g}} \{\frac{1}{2}(U_{a}U^{a} +
U^{a}U_{a})
+ \frac{1}{2}(K_{+}K^{+} + K^{+}K_{+}) + K_{3}^{2} \\
     &   & - \frac{k+1}{2}(\psi_{a} \partial \psi^{a} +
\psi^{a} \partial \psi_{a}) +H_{a}^{b} \psi^{a} \psi_{b} +
\frac{1}{4}\Omega_{ab}
 \psi^{a} \psi^{b} \Omega^{cd} \psi_{c} \psi_{d} \}(z)
\end{array}
\end{equation}

The generators of the two $SU(2)$  currents take the form
\begin{equation}
\begin{array}{lll}

V_{3}^{+}(z) & = & K_3(z) +
\frac{1}{2}( \psi_{+} \psi^{+} + \chi_{+} \chi^{+})(z) \\
V_{+}^{+}(z) & = & (V_{1}^{+}+ i V_{2}^{+})(z) =
(K_{+} - \psi_{+} \chi_{+})(z) \\
V_{-}^{+}(z) & = &(V_{1}^{+}- i V_{2}^{+})(z) =
(K^{+} + \psi^{+} \chi^{+})(z) \\
V_{3}^{-}(z) & = & \frac{1}{2}(\psi^{a}\psi_{a} +
\psi^{+}\psi_{+} + \chi_{+}\chi^{+} )(z) \\
V_{+}^{-}(z) & = & (V_{1}^{-}+ i V_{2}^{-})(z) =
(\psi^{+} \chi_{+} -
 \frac{1}{2}\Omega_{ab} \psi^{a} \psi^{b})(z)   \\
V_{-}^{-}(z) & = & (V_{1}^{-}- i V_{2}^{-})(z) =
(\chi^{+} \psi_{+} -
 \frac{1}{2}\Omega^{ab} \psi_{a} \psi_{b})(z)

\end{array}
\end{equation}
The  U(1) current of the $N=4$ SCA is  $Z(z)$ and the  four
dimension $\frac{1}{2}$ generators  are simply the
fermion fields $\psi_{+}(z), \psi^{+}(z), \chi_{+}(z)$ and $\chi^{+}(z)$.
One finds that the levels of the two $SU(2)$ currents are $k^{+}= k+1$ and
$k^{-}= \check{g} - 1$. The central charge of the $N=4$ SCA turns out to be
$c=\frac{6k^+k^-}{k^++k^-}$.
 The above realization of the $N=4$ SCA corresponds to the coset
space $G \times U(1)/H$.

By decoupling the four dimension $\frac{1}{2}$ generators and the $U(1)$
current $Z(z)$ one obtains a non-linear $N=4$ SCA \cite{GS,GPTV} a la
Bershadsky and Knizhnik \cite{VK,MB}.
The realization given above for the $N=4$ SCA leads to very simple expressions
for the
supersymmetry generators  of the non-linear $N=4$ algebra \cite{MG93b}
\eqn
\tilde{G}^{+} \equiv \frac{1}{\sqrt{2}} (\tilde{G}_{1}+i\tilde{G}_{2})& =&
\sqrt{\frac{2}{k+\check{g}}} U_{a} \psi^{a} \nn
\tilde{G}^{-} \equiv \frac{1}{\sqrt{2}} (\tilde{G}_{1}-i\tilde{G}_{2})& =&
\sqrt{\frac{2}{k+\check{g}}} U^{a} \psi_{a} \nn
\tilde{G}^{+K} \equiv \frac{1}{\sqrt{2}} (\tilde{G}_{3}+i\tilde{G}_{4})&=&
\sqrt{\frac{2}{k+\check{g}}} \Omega^{ab} U_{a} \psi_{b}   \nn
\tilde{G}^{-K} \equiv \frac{1}{\sqrt{2}} (\tilde{G}_{3}-i\tilde{G}_{4})& =&
\sqrt{\frac{2}{k+\check{g}}} \Omega_{ba} U^{a} \psi^{b}
\enn
with the central charge $\tilde{c} = c - 3$.

It is clear from the above expressions for the generators that the non-linear
$N=4$  SCA is
realized over the symmetric space
\eq
G/H\times SU(2)
\en
which is the unique quaternionic symmetric space associated with $G$
\cite{AV89} .

\section{$N = 4$ Supersymmetric Gauged WZW Models}
\setcounter{equation}{0}
So far we have been discussing the construction of extended
\sca s and study of their chiral rings using algebraic methods. In this section
we shall study a certain class of Lagrangian field theories
, namely supersymmetric gauged WZW models,  that are invariant
under extended superconformal groups. Gauged WZW models were studied in
\cite{GWZW,EW92b}.
The $N=1$ supersymmetric
ordinary WZW models were studied in \cite{RR,DKPR,SSTP1} and  their gauged
versions in \cite{HS2}. $N=2$ supersymmetric gauged WZW models were
studied by Witten \cite{EW92}. Witten's results were extended to
  $N=4$ gauged
WZW models in \cite{MG93b}.

The WZW action at level $k$ is given by $kI(g)$ where
\eq
I(g) = -\frac{1}{8\pi} \int_{\Sigma} d^2 \sigma \sqrt{h} h^{ij}
Tr(g^{-1}\partial_ig \cdot g^{-1}
\partial_j g) -i \Gamma
\en
with the WZ functional \cite{WZ} given by \cite{EW84}
\eq
\Gamma = \frac{1}{12\pi} \int_M d^3 \sigma \epsilon^{ijk} Tr ( g^{-1}
\partial_ig \cdot
g^{-1} \partial_jg \cdot g^{-1} \partial_k g)
\en
 $M$ is a three manifold whose boundary is the Riemann surface $\Sigma$
with metric $h$. We choose the  metric $h_{z\bar{z}} =h^{z\bar{z}}=1$ and
work with complex coordinates $z$ and $\bar{z}$. $g$ represents the group
element that maps $\Sigma$ into the group $G$.
The supersymmetric WZW action  $I(g,\Psi)$  is obtained by adding to $I(g)$
the free action of Weyl fermions $\Psi_{L}$ and $\Psi_{R}$ in the
complexification of the
adjoint representation of $G$
\cite{EW92}:\footnote{We shall consider only models that have equal
number of supersymmetries in both the left and the right moving sectors.}
\eq
I(g,\Psi) =I(g) + \frac{i}{4\pi} \int d^2z Tr(\Psi_L \partial_{\bar{z}} \Psi_L
+
\Psi_{R} \partial_z \Psi_R )
\en
It is invariant under the supersymmetry transformations
\eqn
\delta g& =& i \epsilon_- g \Psi_L +i \epsilon_+ \Psi_R g  \nn
\delta \Psi_L& =& \epsilon_- (g^{-1} \partial_z g -i \Psi_{L}^{2})  \nn
\delta \Psi_R& =& \epsilon_+ (\partial_{\bar{z}} g g^{-1} + i \Psi_{R}^{2} )
\enn

Gauging any diagonal subgroup $H$ of the $G_L\times G_R$
 symmetry of the WZW model
leads to an anomaly free theory.
 The gauge invariant action, which does not involve any kinetic
energy term for the gauge fields,  can be written as:
\eq
I(g,A) = I(g) + \frac{1}{2\pi} \int_{\Sigma} d^2z Tr(A_{\bar{z}} g^{-1}
\partial_z g - A_z \partial_{\bar{z}}g g^{-1} +A_{\bar{z}}g^{-1} A_{z}g
-A_{\bar{z}}
A_z)
\en
where $A_{z}, A_{\bar{z}}$ are the matrix valued gauge fields belonging
to the subgroup $H$.
It is invariant under the gauge transformations \cite{EW92}:
\eqn
\delta g& =& [u,g] \nn
	\delta A_i& =& -D_iu = -\partial u -[A_i,u]
\enn
Denoting as ${\cal G}$ and ${\cal H}$
 the complexifications the Lie algebras of
 $G$ and $H$ one has the orthogonal decomposition
\eq
{\cal G} = {\cal H} \oplus {\cal T}
\en
where ${\cal T}$ is the orthocomplement of ${\cal H}$.
 To supersymmetrize the  WZW model with the gauged subgroup $H$
 one introduces Weyl fermions with
values in ${\cal T}$ minimally coupled to the gauge fields and otherwise free
\eq
I(g,A,\Psi) = I(g,A) + \frac{i}{4\pi} \int d^2z Tr(\Psi_L D_{\bar{z}} \Psi_L
+ \Psi_R D_z \Psi_R )
\en
This action is invariant under the supersymmetry transformation laws:
\eqn
\delta g& =& i \epsilon_- g \Psi_L +i \epsilon_+ \Psi_R g \nn
\delta \Psi_L& =& \epsilon_- (1-\Pi) (g^{-1} D_z g -i \Psi_L^2 ) \nn
\delta \Psi_R& =& \epsilon_+ (1- \Pi) (D_{\bar{z}} g g^{-1} +i \Psi_R^2)  \nn
\delta A& =& 0
\enn
where $\Pi$ is the orthogonal projection of ${\cal G}$ onto ${\cal H}$.
As shown by Witten  if the gauge group $H$ is such that
 the coset space $G/H$ is Kahler the above action has
$N=2$ supersymmetry\cite{EW92}. For $G/H$  Kahler the subspace
 ${\cal T}$ can be decomposed as
\eq
{\cal T} = {\cal T}_+ \oplus {\cal T}_-
\en
where ${\cal T}_+$ and ${\cal T}_-$ are in complex conjugate representations of
$H$.
The action can then be written in the form \cite{EW92}
\eq
I(g,\Psi,A) = I(g,A) + \frac{i}{2\pi} \int d^2z Tr ( \beta_L D_{\bar{z}}
\alpha_L + \beta_R D_z \alpha_R )
\en
where
\eqn
\alpha_L& =& \Pi_+ \Psi_L  \nn
 \beta_L& =& \Pi_- \Psi_L \nn
\alpha_R& =& \Pi_+ \Psi_R \nn
\beta_R& =& \Pi_- \Psi_R
\enn
with $\Pi_+$ and $\Pi_-$ representing the projectors onto the subspaces
${\cal T}_+$ and ${\cal T}_-$.
Denoting the chiral and anti-chiral supersymmetry generators in left  and right
moving
sectors as $G_L , \bar{G}_L$ and $G_R , \bar{G}_R$ , respectively,
one finds \cite{EW92} that they satisfy the $N=2$ supersymmetry algebra

\eqn
\{G_L,\bar{G}_L\}&=&-iD_z \nn
\{G_R,\bar{G}_R\}&=&-iD_{\bar{z}}  \label{eq:AR}
\enn
The  gauged WZW models are known to be conformally invariant. Hence the
 $N=2$ global supersymmetry of these theories implies $N=2$ superconformal
invariance.

 Since the existence of a second supersymmetry in a supersymmetric
gauged WZW model is guaranteed by the K\"{a}hlerian property of
 the coset space $G/H$  , i.e
when it
admits a complex structure,
 to have $N=4$ supersymmetry one needs
 coset spaces with three complex structures
 which anti-commute with each other
and form a closed algebra. Therefore one expects
supersymmetric WZW models based on the groups $G\times U(1)$ of the previous
section with
a gauged subgroup $H$ such that $G/H\times SU(2)$ is a quaternionic symmetric
 space to
 actually have $N=4$ supersymmetry. Let us show that this is indeed the case
\cite{MG93b}.

We shall designate the generators of $G\times U(1)$ as we did in previous
sections
,i.e $K_0, K_1,K_2,K_3,U_a,U^a$ and $H^b_a$ , where $K_0$ is the generator of
the  additional
$U(1)$ factor normalized such that
\eq
Tr K_0^2 =Tr K_1^2=Tr K_2^2 =Tr K_3^2
\en
 The fermions associated with the grade $\pm1$ subspaces of the Lie
algebra of $G$ will be denoted as $\psi_a,\psi^a$ as before and the fermions
associated with $K_0$ and $K_i$
will be denoted as $\xi^0, \xi^i (i=1,2,3)$.Thus  the fermions
in the coset $G\times U(1)/H$ can be represented as
\eq
\Psi = 2K_0 \xi^0 + 2K_i \xi^i + U_a \psi^a + U^a \psi_a
\en
for both the left and the right moving sectors.
The coset $G\times U(1)/H$ can be given a Kahler decomposition such that

\eq
\Psi= \A + \B
\en
where
\eqn
\A& =& U_a \psi^a + K_+ (\xi^1-i\xi^2) + (K_3+iK_0)(\xi^3-i\xi^0) \nn
\B& =& U^a \psi_a + K_- (\xi^1 +i \xi^2) + (K_3-iK_0)(\xi^3+i\xi^0)
\enn
The complex structure $C_3$ corresponding to this
Kahler decomposition acts on $\Psi$ as
\eq
C_3 \Psi = -i \A +i\B
\en
where the index 3 in $C_3$ signifies the fact in the subspace
$G/H \times SU(2)$ its action corresponds to commutation with
the generator $-iK_3$.
One can similarly give a Kahler decomposition of
 the coset space $G\times U(1)/H$
which selects out $K_1$ or $K_2$. For the decomposition with respect
to $K_1$ we have:
\eqn
\Psi& =& \A^{\prime} + \B^{\prime} \nn
\A^{\prime}& =& U_a^{\prime} \psi^{a^{\prime}}
 +(K_2+iK_3)(\xi^2-i\xi^3) + (K_1+iK_0)(\xi^1-i\xi^0)  \nn
\B^{\prime}& =& U^{a^{\prime}} \psi_a^{\prime} +(K_2-iK_3)(\xi^2+i\xi^3)
+ (K_1-iK_0)(\xi^1+i\xi^0)
\enn
Under the action of the corresponding complex structure $C_1$ we have
\eq
C_1 \Psi = -i\A^{\prime} +i\B^{\prime}
\en
In the case of $K_2$ one finds
\eqn
\Psi&=& \A^{\prime \prime} +\BPP \nn
\APP& =&  U_a^{\prime \prime}\psi^{a^{\prime \prime}}
 +(K_3+iK_1)(\xi^3-i\xi^1) + (K_2+iK_0)(\xi^2-i\xi^0) \nn
\BPP& =& U^{a^{\prime \prime}} \psi_a^{\prime \prime}
+ (K_3-iK_1)(\xi^3+i\xi^1) +(K_2-iK_0)(\xi^2+i\xi^0)
\enn
with the complex structure action
\eq
C_2 \Psi = -i \APP +i\BPP
\en
The fermionic part of the action
$I(g, \Psi, A)$ can then be written in three different
ways involving the pairs $(\A,\B),(\AP,\BP)$ and $(\APP,\BPP)$.
\eqn
I(\Psi,A)&=& \frac{i}{4\pi}\int d^2z Tr(\Psi_L D_{\bar{z}}\Psi_L
+ \Psi_R D_z \Psi_R ) \nn
&=& \frac{i}{2\pi}\int d^2z Tr( \B_L D_{\bar{z}}\A_L +
\B_R D_z \A_R) \nn
&=& \frac{i}{2\pi}\int d^2z Tr(\BP_L D_{\bar{z}} \AP_L +
\BP_R D_z \AP_R) \nn
&=& \frac{i}{2\pi} \int d^2z Tr(\BPP_L D_{\bar{z}} \APP_L +
\BPP_R D_z \APP_R)
\enn
     For each form of the
action in terms of an $(\A,\B)$ pair one can define a pair of supersymmetry
transformations in each sector. Let us denote them as $(G,\bar{G})
,(G^{\prime}.\bar{G}^{\prime})$ and $(G^{\prime \prime},\bar{G}^{\prime
\prime})$:
\eqn
(G,\bar{G}) &\leftrightarrow & (\A,\B) \nn
(G^{\prime}, \bar{G}^{\prime}) & \leftrightarrow &
(\AP, \BP) \nn
(G^{\prime \prime}, \bar{G}^{\prime \prime}) & \leftrightarrow &
(\APP,\BPP)
\enn
Each pair of these operators in both sectors satisfy the $N=2$ supersymmetry
algebra given by the equations \ref{eq:AR}.   However they are not all
independent. The sum of each pair gives the manifest $N=1$ supersymmetry
generator
of the model in both sectors, which we shall denote as $G^0$:
\eq
G^0 =\frac{1}{\sqrt{2}}(G+\bar{G}) = \frac{1}{\sqrt{2}}(G^{\prime}+
\bar{G}^{\prime})= \frac{1}{\sqrt{2}}(G^{\prime \prime}+ \bar{G}^{\prime
 \prime})
\en
They satisfy
\eqn
\{G^0_L,G^0_L\}&=& -iD_z \nn
\{G^0_R,G^0_R\}&=&-iD_{\bar{z}}
\enn
One then has three additional supersymmetry generators (in each sector)
\eqn
G^3 &=& \frac{1}{i\sqrt{2}}(G-\bar{G}) \nn
G^1 &=& \frac{1}{i\sqrt{2}}(G^{\prime}-\bar{G}^{\prime}) \nn
G^2 &=& \frac{1}{i\sqrt{2}}(G^{\prime \prime}
-\bar{G}^{\prime \prime})
\enn
 Each one of these three supersymmetry generators anticommute with $G^0$.
To prove that $G^{\mu} (\mu=0,1,2,3)$
form an $N=4$ superalgebra we need to
further show that the $G^i (i=1,2,3)$ anticommute with each other.
To prove this we first note that the complex structures
$C_i$ obey the relation
\eq
C_i C_j = C_k
\en
where $i,j,k$ are in cyclic permutations of $(1,2,3)$ and the
action is invariant under the replacement of $\Psi$ by $C_i \Psi$.
If we start from an action with $\Psi$ replaced by ,say, $C_1 \Psi $
then the
manifest $N=1$ supersymmetry will be generated by $G^1$ and the second
supersymmetry generated by the Kahler decomposition with respect to the
complex structure $C_3$ will be $G^2$ since $C_3 C_1=C_2$. Hence by the
results of Witten on $N=2$ supersymmetric gauged WZW models we have
\eq
\{G^1,G^2\}=0
\en
and by cyclic permutation we find
\eq
\{G^2,G^3\}=\{G^3,G^1\}=0
\en\
Thus the four supersymmetry generators
$G^{\mu}$ satisfy the $N=4$ supersymmetry algebra:
\eqn
\{ G^{\mu}_L, G^{\nu}_L\}& =& -i \delta^{\mu \nu} D_z \nn
\{ G^{\mu}_R , G^{\nu}_R\}& = & -i \delta^{\mu \nu} D_{\bar{z}}\nn
\mu,\nu,...&=&0,1,2,3
\enn
Since the gauged $WZW$ models considered above
are known to be conformally invariant we have
thus proven that they are invariant under $N=4$ superconformal
transformations.

\  \\

{\it Acknowledgements:} This talk was written up during my stay at the
Institute for Theoretical Physics of the University of Helsinki.
I would like to thank Antti Niemi and other members of the Institute for
their kind hospitality.


\end{document}